\documentclass[12pt]{iopart}
\usepackage{amssymb}
\usepackage{iopams,bm,verbatim}
\usepackage[T1]{fontenc}
\usepackage{subeqn}
\def\Journal#1#2#3#4{{#4} {\it #1} {\bf #2}, #3 }

\def\Xc{\bar{X}}
\def\td{\tau'}
\def\tdbar{\overline{\tau}'}
\def\tbar{\overline{\tau}}
\def\rd{\rho'}
\def\rdbar{\overline{\rho}'}
\def\phidbar{\overline{\phi}'}
\def\phibar{\overline{\phi}}
\newcommand{\w}[1]{\bm{#1}} 

\def\tho{\textrm{\TH}}
\def\thd{\tho '}
\def\et{\eth}
\def\etd{\eth '}
\def\ud{\textrm{d}}
\newcommand{\be}{\begin{equation}}
\newcommand{\ee}{\end{equation}}

\begin{document}
%

\title{
Killing spinor space-times and constant-eigenvalue Killing tensors}

\author{D.~Beke
\footnote[3]{To
whom correspondence should be addressed (david.beke@ugent.be)}, N.~Van den Bergh, L.~Wylleman\dag}

\address{Ghent University, Department of Mathematical Analysis IW16, \\ Galglaan 2, 9000 Ghent, Belgium}
\address{\dag\ Utrecht University, Department of Mathematics, \\ Budapestlaan 6, 3584 CD Utrecht, the Netherlands.}

\begin{abstract}
      A class of Petrov type D Killing spinor space-times is presented, having the peculiar property that their conformal representants can only admit Killing tensors with constant eigenvalues.
\end{abstract}

\pacs{04.20.Jb}

\section{Introduction}
In a recent paper~\cite{KS1_2} ``KS space-times'' were defined as non-conformally flat space-times admitting a \emph{non-null} valence
two spinor $X_{AB}$, satisfying the conformally invariant Killing spinor equation
\begin{equation}\label{eqKS}
\nabla _{A'(A}X_{BC)}=0 .
\end{equation}
The associated two-form $D_{ab}= X_{A B} \epsilon_{A'B'}+\Xc_{A'B'} \epsilon_{AB}$ is a conformal Killing-Yano (CKY)
tensor and, being non-null, KS space-times form the subclass of Petrov type D CKY space-times. Their repeated principal Weyl spinors
are aligned with the principal spinors of $X_{AB}$ \cite{PlebHacyan76} and define geodesic shear-free null congruences.

The square $P_{ab}={D_a}^c D_{cb}$ of a CKY tensor $D$ is a conformal Killing tensor of Segre type $[(11)(11)]$ and hence Killing spinors allow the construction
of constants of motion along null-geodesics; they have the additional significance \cite{KamranMcL1984} that they are the geometric objects
from which one may construct symmetry operators for the massless Dirac equation. KS space-times necessarily include all space-times which are
conformally related to  Petrov type D Killing-Yano space-times: the inclusion is strict, a counter-example being given by the
Kinnersley Case III metrics \cite{CzaporMcLen1982,DebeverMcLen1981}.

A KS space-time always admits a conformal representant in which the trace of the associated conformal Killing tensor $P_{ab}$ is constant.
In this representant the conformal Killing tensor becomes a Killing tensor, but has two constant eigenvalues.
One may ask whether different conformal representants exist,
admitting Killing tensors with non-constant eigenvalues: in the affirmative case
this greatly simplifies the construction of the canonical line-elements of KS space-times (see \cite{Jeffryes1984} for the case where the
two eigenvalues provide independent functions of the coordinates and \cite{McLenVdB1993, KS1_2} for the case where one of the eigenvalues is
constant). In this paper we present a first example of KS space-times in which any associated Killing tensor always has two constant eigenvalues.

As in \cite{KS1_2} the Geroch-Held-Penrose formalism~\cite{GHP} is used, but to ease comparison with the literature (particularly with regard to
a possible interpretation of the energy-momentum tensor) we follow the notation and
sign conventions of \cite{Kramer}: the tetrad basis vectors are taken as $\w{k},\w{\ell},\w{m},\overline{\w{m}}$ with $- k^a \ell_a = 1 = m^a
\overline{m}_a$. The correspondence with the Newman-Penrose operators and the basis one-forms
is taken as $(m^a,\overline{m}^a, \ell^a, k^a) \leftrightarrow (\delta, \overline{\delta}, \Delta, D)$ and
$(\overline{m}_a,m_a,-k_a,-l_a) \leftrightarrow
(\w{\omega}^1,\w{\omega}^2,\w{\omega}^3,\w{\omega}^4)$ (this has the effect of changing the sign of the trace $K$ in \cite{KS1_2}). For completeness we
repeat in section 2 the construction of the main equations, omitting however all details.

\section{Preliminaries}
Writing the Killing spinor as $X_{AB}=X o_{(A} \iota_{B)}$,
the components of (\ref{eqKS}) imply
\begin{equation}\label{eqKSa}
\kappa=\sigma=0
\end{equation}
and
\begin{eqnarray}
\tho X = -\rho X, \label{KSb1}\\
\et X = -\tau X, \label{KSb2}
\end{eqnarray}
together with their `primed versions' and $X'=X$.
It follows that $\Psi_2$ is the only non-vanishing component of the Weyl spinor; the spin coefficients $\rho$, $\rho'$, $\tau$ and
$\tau'$ are assumed to be non-zero (otherwise we re-obtain the metrics of \cite{Jeffryes1984, McLenVdB1993}). A conformal representant $(\mathcal{M},\hat{g})$ is fixed by imposing $|\hat{X}|=1$. In the manifold
$(\mathcal{M},\hat{g})$ one has $\hat{\rho}+\overline{\hat{\rho}}=\hat{\rho}'+\overline{\hat{\rho}}'=
\hat{\tau}+\overline{\hat{\tau}}'=0$,
while in the $(\mathcal{M},g)$ manifold with $g=
\Omega^2\hat{g}$ one has $\Omega^2=X\overline{X}$.

Defining the trace-free conformal Killing tensor
\be
P_{ab}=X_{AB}\overline{X}_{A'B'}=\frac{\Omega^2}{2}(m_{(a} \overline{m}_{b)}+\ell_{(a}n_{b)}),
\ee
one can show that $P_{ab}$ will be the trace-free part of a Killing tensor $K_{ab}$ if
\be
\tho K + \Omega^2(\rho+\overline{\rho})=0=
\et K -\Omega^2(\tau+\tdbar) \label{KTcomps},
\ee
or, 
in terms of the eigenvalues $a=(\Omega^2+K)/4$ and $b=(\Omega^2-K)/4$ of
\be
K_{cd}= 2 a m_{(c}\overline{m}_{d)}+2 b \ell_{(c} n_{d)} :
\ee
\be
\tho b =0 = \et a \label{eqeta}.
\ee
By (\ref{KSb1},\ref{KSb2}) and $\Omega^2=X \overline{X}$ this implies
\begin{eqnarray}
\tho a = -(a+b) (\rho+\overline{\rho}), \label{KTextra}\\
\et b = -(a+b) (\tau+\tdbar), \label{KTextrb}
\end{eqnarray}
such that (\ref{eqeta},\ref{KTextra},\ref{KTextrb}) can alternatively be written as
\begin{eqnarray}
\mathrm{d} a = -(a+b) [(\rho+\overline{\rho})\w{\omega}^4+(\rho'+\overline{\rho}')\w{\omega}^3] \label{dmaina} \\
\mathrm{d} b = -(a+b)[(\tau+\tdbar)\w{\omega}^1 + (\overline{\tau}+\td)\w{\omega}^2] \label{dmainb}.
\end{eqnarray}
KS space-times therefore admit at least one conformal representant, $(\mathcal{M},\hat{g})$, in which a Killing tensor exists, which however has constant eigenvalues.
Insisting on the existence of a conformal representant in which the eigenvalues are
\emph{not} both constants, extra integrability conditions result from the equations $\mathrm{d}\mathrm{d} a= \mathrm{d}\mathrm{d} b=0$.

It is preferable to manipulate all ensuing equations in the $(\mathcal{M},\hat{g})$ manifold, where we
drop the $\hat{}$ symbol from here onward: the remaining spin coefficients are then $\rho=-\overline{\rho}$, $\rd=-\rdbar$ and $\tau=-\tdbar$
and the integrability
conditions for the system (\ref{KSb1},\ref{KSb2}) simplify to
\begin{eqnarray}\label{specKS}
\thd \rho -\tho \rd =0 , \label{specKSa}\\
\et \td -\etd \tau =0, \label{specKSb} \\
\tho \td -\etd \rho = 0 .\label{specKSc}
\end{eqnarray}
The GHP equations reduce then to the system
\begin{eqnarray}
\tho \rho =0, \label{thrho}\\
\et \rho = 2 \rho \tau +\Phi_{01},\label{etrho}
\end{eqnarray}
\begin{eqnarray}
\tho \tau = 2 \rho \tau + \Phi_{01}, \label{thtau} \\
\et \tau = 0, \label{ettau}
\end{eqnarray}
\be
\tho \rd -\et \td
= -\rho \rd -\tau \tbar-\Psi_2-\frac{1}{12} R \label{thrd_ettd}\\
\ee
and impose the following restrictions on the curvature:
\begin{eqnarray}
\label{defPhi00}\Phi_{00}=-\rho^2,
 \Phi_{02}=-\tau^2,\\
 E=-\frac{R}{12}-\rho \rd +\tau \td,
\end{eqnarray}
where $E$ is the real part of $\Psi_2=E+i H$.

Introducing 0-weighted quantities $u,v$ (both real and with $u'=u$, $v'=v$) and $\phi,\phi'$ (complex) by
\begin{eqnarray}
R=8(u-v)-16\rho \rd, \\
\Phi_{11}=u+v-2 \rho \rd,\\
\label{defPhi01}\Phi_{01}=-3 \rho \tau -2\frac{\rho}{\td} \phi ,
\end{eqnarray}
one can show~\cite{KS1_2} that a conformal representant admitting a Killing tensor with non-constant eigenvalues can only exist when one of the following conditions hold\footnote{note the print error in equation (40) of \cite{KS1_2}}:
\begin{eqnarray}
\exists \textrm{ non-constant } b \Longleftrightarrow \phi+\overline{\phi} = \phi'+\phidbar = 0, \label{condKS1}\\
\exists \textrm{ non-constant } a \Longleftrightarrow \phi+\phidbar=0 . \label{condKS2}
\end{eqnarray}
The corresponding space-times were called $KS_1$ or $KS_2$ respectively. All $KS_1 \cap KS_2$ space-times were discussed in \cite{Jeffryes1984}, while
the remaining space-times belonging to $KS_1 \cup KS_2$ were dealt with in \cite{McLenVdB1993,KS1_2}. It was left undecided however whether KS space-times existed which didn't belong to $KS_1 \cup KS_2$.
In the next paragraph we show that the answer is affirmative: KS space-times do exist in which both conditions (\ref{condKS1},\ref{condKS2}) are violated.

Introducing 0-weighted extension variables $U$ (real) and $V$ (complex), by
\begin{eqnarray}
\tho \rd = \thd \rho = -i U , \label{eerste}\\
\thd \phi = \rd( 2\frac{\phi \phi'}{|\tau|^2}+V)+i \frac{U\phi}{\rho},
\end{eqnarray}
the Bianchi identities and the `first level' integrability conditions on $\rho,\rd,\tau,\td$ provide expressions for the directional derivatives of $\phi, u, v, U, V$ and $H$.
 Constructing the `second level' integrability conditions, by applying the commutator operators to the latter variables, leads to an over-determined system of equations, the general solution of which so far has not been obtained. The system can be greatly simplified however by assuming that $\phi'-\phi$ is real: defining 0-weighted real variables $r > 0$ and $m > 0$ by
\be
r^2=Q\rho\rd,\ m=|\tau|\ \ (Q=\pm 1) \label{rmdef},
\ee
one can show that this implies
\begin{eqnarray}
\phi' = \phi,\\
V = \frac{2}{m^2} (m^4-2\phi \phibar),\\
U = 2 i Q \frac{r}{m}^2(\phi-\phibar),\\
H = \frac{2 i}{m^2} (\phi -\phibar)(Q r^2+m^2),\\
u = \frac{1}{2m^2} (3 m^4+6 Q r^2 m^2-2 (\phi+\phibar)Q r^2 -4 \phi \phibar ),\\
v = -\frac{1}{2m^4}(2 m^6+3 Q r^2 m^4-2(\phi+\phibar)m^4-4 Q r^2 \phi \phibar).
\end{eqnarray}

In terms of the curvature components this gives
\begin{eqnarray}
\label{defR}R = -\frac{4}{m^2}[(4\phi\phibar+2m^2(\phi+\phibar)-5m^4)(Qr^2+m^2)],\\
\label{defPsi2}\Psi_2 = \frac{4}{3m^4}(\phi+2m^2)(\phibar-m^2)(Qr^2+m^2),\\
\Phi_{11} = \frac{1}{2m^4}(-4\phi\phibar+2(\phi+\phibar)m^2+m^4)(m^2-Qr^2).
\end{eqnarray}

Herewith the differential equations for the remaining 0-weighted variables can be succinctly written as
\begin{eqnarray}
\ud X = X (-\tau \w{\omega}^1+m^2 \tau^{-1} \w{\omega}^2 -Qr^2\rho^{-1} \w{\omega}^3 -\rho \w{\omega}^4),\\
\ud r = \frac{r(\phi-\phibar)}{m^2 X} \ud X,\\
\ud m = \frac{\phi-\phibar}{m X} \ud X,\\
\ud \phi = \frac{2(\phi\phibar-m^4)}{m^2 X} \ud X, \label{dphi}
\end{eqnarray}
while  the GHP-derivatives of the weighted quantities $\rho$ and $\tau$ are given by
\begin{eqnarray*}
\et \rho = \frac{\rho \tau}{m^2} (2 \phi-m^2),\quad \etd\rho=\frac{\rho}{\tau}(2 \phibar -m^2), \thd\rho=\frac{2Qr^2}{m^2}(\phi-\phibar),\quad \tho\rho=0,\\
\et\tau=0,\quad \etd\tau=2(\phibar-\phi),\quad\thd\tau=\frac{Qr^2\tau}{m^2\rho}(m^2-2\phibar),\quad \tho\tau=\frac{\rho \tau}{m^2}(2\phi-m^2).
\end{eqnarray*}
One can easily verify that the integrability conditions for this system are identically satisfied, such that corresponding solutions exist, and that $m/r$ and $\Re{\phi}$ are constants, with $\Re{\phi}\neq 0$ as otherwise both conditions (\ref{condKS1},\ref{condKS2}) would hold. Since $\rho\tau\neq0$ the tetrad can be invariantly fixed (up to interchange of $\w{k}$ and $\w{l}$), and one invokes  from (\ref{defPhi00}), (\ref{defPhi01}) and (\ref{defR}-\ref{dphi})
that the components of the Riemann tensor and its covariant derivatives wrt such a tetrad contain at most one functionally independent function. Thus~\cite{Karlhede,Kramer} the corresponding space-times admit a 3- or 4-dimensional maximal group of isometries. The first possibility corresponds to non-constant $r$ and will be discussed elsewhere. In the next paragraph we will discuss the solutions corresponding to the homogeneous, purely electric case
\be
\Psi_2\in \mathbb{R} \Leftrightarrow \phi\in \mathbb{R} \Leftrightarrow r \textrm{ and } m \textrm { constant.}
\ee
By (\ref{dphi}) this furthermore implies that $\phi = \pm m^2$, where $\phi=m^2$ leads to conformally flat solutions, cf.\ (\ref{defPsi2}). For $\phi=-m^2$, however, we get
\begin{eqnarray}
\Psi_2= -\frac{8}{3}(m^2+Qr^2),\qquad R=20(m^2+Qr^2).\label{KS3const}
\end{eqnarray}
Notice that when $Q=-1$ we obtain conformally flat solutions (with $R=0$) if $m=r$, and we will therefore restrict to $m\neq r$ in this case.

\section{Purely electric solutions}
In order to integrate the above system under conditions (\ref{KS3const}), we switch to the Newman-Penrose formalism\cite{NP} and fix a boost and rotation such that $\tau=m$ and $\rho=iQr$. For the spin coefficients $\alpha, \beta, \epsilon, \gamma$ this implies
\be
\epsilon=\frac{3}{2}iQr,\quad \gamma=\frac{3}{2}ir,\quad\alpha=\beta=\frac{3}{2}m\label{spincoeffs}
\ee
Introducing new basis one-forms by
\be
\fl \w{\Omega^1}=i (\w{\omega}^1-\w{\omega}^2),\ \w{\Omega}^2=\w{\omega}^1+\w{\omega}^2,\ \w{\Omega}^3=\w{\omega}^3+Q\w{\omega}^4,\ \w{\Omega}^4=\w{\omega}^3-Q\w{\omega}^4,
\ee
such that the line-element reads
\be
\ud s^2 = 2({\w{\Omega}^1}^2+{\w{\Omega}^2}^2-Q {\w{\Omega}^3}^2+Q{\w{\Omega}^4}^2),\label{linelementOm}
\ee
the Cartan equations become
\begin{eqnarray}
\ud \w{\Omega}^1 = 2 r \w{\Omega}^3 \wedge \w{\Omega}^2,\nonumber \\
\ud \w{\Omega}^2 = -2 \w{\Omega}^3 \wedge (r \w{\Omega}^1 + m Q \w{\Omega}^4),\nonumber\\
\ud \w{\Omega}^3 = 2 Q \w{\Omega}^2 \wedge (r \w{\Omega}^1 + m Q \w{\Omega}^4),\nonumber \\
\ud \w{\Omega}^4 = -2 m \w{\Omega}^3 \wedge \w{\Omega}^2.\label{Cartan}
\end{eqnarray}
It follows that $m\w{\Omega}^1+r\w{\Omega}^4$ is exact and that (the dual vectorfield of) $\sqrt{-Q}\w{\Omega}^2+\w{\Omega}^3$ is hypersurface-orthogonal.
\subsection{$Q=+1$}
When $Q=1$ one immediately can introduce coordinates $t,x$ and $y$ such that
\begin{eqnarray}
\w{\Omega}^1 = \ud t -\frac{r}{m} \w{\Omega}^4,\\
\w{\Omega}^3+i \w{\Omega}^2 = e^{u+iv}\ud (x+i y).
\end{eqnarray}
The cases $m=r$ and $m\neq r$ must now be treated separately. When $m\neq r$ equations (\ref{Cartan}) integrate to
\be
\w{\Omega}^4= \frac{m}{2(m^2-r^2)}(\ud v-2 r \ud t+u_{y} \ud x -u_{x} \ud y),
\ee
with $u(x,y)$ a solution of the Liouville equation
\be
u_{xx}+u_{yy}=4(m^2-r^2) e^{2 u}. \label{Liou}
\ee
This equation can be solved analytically, yielding solutions involving a free analytic function $F(x+iy)$, but this result is not needed here. Since all spin coefficients are constant and independent of $F$, all components of the Riemann tensor with respect to the fixed tetrad are constant and independent of $F$. This implies \cite{Karlhede} that line elements involving different $F$ are equivalent, and we can choose the particular solution
\begin{equation}
e^{u}=\frac{1}{\mathcal{K}(x,y)}, \quad \mathcal{K}(x,y)=1-\left(m^2-r^2\right)\left(x^2+y^2\right).
\end{equation}
Hence the line element is given by (\ref{linelementOm}), where
 \begin{eqnarray}
 \w{\Omega}^1 = \frac{1}{2(m^2-r^2)}(2 m^2 \ud t - r \ud v) -\frac{r}{\mathcal{K}(x,y)}(y\ud x-x \ud y),\nonumber \\
 \w{\Omega}^2 = \frac{1}{\mathcal{K}(x,y)}(\sin v \ud x+\cos v \ud y ),\nonumber\\
 \w{\Omega}^3 = \frac{1}{\mathcal{K}(x,y)}(\cos v \ud x -\sin v \ud y) ,\nonumber\\
 \w{\Omega}^4 = \frac{m}{2(m^2-r^2)}(-2 r \ud t +\ud v) + \frac{m}{\mathcal{K}(x,y)}(y\ud x-x \ud y) \label{posQcaseA}.
 \end{eqnarray}

When $m=r$ the integration of the Cartan equations is straightforward and leads to the line element (\ref{linelementOm}) with
 \begin{eqnarray}
 \w{\Omega}^1 = \ud(v-u)+2 m x \,\ud y ,\nonumber\\
 \w{\Omega}^2 = \sin 2mv \,\ud x+\cos 2mv \,\ud y,\nonumber\\
 \w{\Omega}^3 = \cos 2mv \,\ud x-\sin 2mv \,\ud y,\nonumber\\
 \w{\Omega}^4 =  \ud u -2 m x \,\ud y.\label{m=r metric}
 \label{posQcaseB}
 \end{eqnarray}
This metric can be obtained as a singular limit of (\ref{posQcaseA}): performing the coordinate transform
\begin{equation}
v\rightarrow 2(m^2-r^2)(v/m-xy)+2rt
\end{equation}
in (\ref{posQcaseA}), taking the limit $r\rightarrow m$ and renaming $(t,v)\rightarrow(v,u)$ one precisely arrives at (\ref{m=r metric}).

\subsection{$Q=-1$ ($m\neq r$)}
It is convenient now to define coordinates $t,x$ and $y$ by
\begin{eqnarray}
\w{\Omega}^1 = \frac{r}{r^2+m^2}(\ud t -\ud y)-\frac{r}{m} \w{\Omega}^4 ,\nonumber\\
 \w{\Omega}^2 = \frac{1}{r^2+m^2}\mathcal{E}^{-1}\ud x -\w{\Omega^3},
\end{eqnarray}
with
\be\label{Ecal}
\mathcal{E}=\exp(2\frac{r^2y+m^2t}{r^2+m^2}).
\ee
The second Cartan equation implies the existence of a new independent function $z$ such that
\be
\w{\Omega}^4=\frac{m}{m^2+r^2}(\ud t + z \ud x),
\ee
after which the remaining Cartan equations integrate to
\be
\w{\Omega}^3 = \frac{1}{2}\mathcal{E}[-\ud z +(z^2+F(x)+\frac{1}{m^2+r^2} \mathcal{E}^{-2})\ud x].
\ee
After a coordinate transformation
\be
y\leftarrow m^2 t+\frac{m^2+r^2}{2} \log y \label{Stefan}
\ee
 the line element becomes (up to a constant re-scaling)
\be
\fl \ud s^2 =-m^2(\ud t+z \ud x)^2+r^2(\ud y+z \ud x)^2+\frac{1}{2}\mathcal{E}^{-2}\ud x^2+\frac{1}{2}(r^2+m^2)^2\mathcal{E}^2[(z^2+F)\ud x-\ud z]^2,
\ee
where $\mathcal{E}$ is still defined by (\ref{Ecal}), in terms of $t$ and the new coordinate $y$.
Under a coordinate transform
\begin{eqnarray}
t\rightarrow t + \xi(x),\quad y\rightarrow y + \xi(x),\quad \ud x \rightarrow e^{2\xi}\ud x,\quad z \rightarrow e^{-2\xi}(z-\xi_x),
\end{eqnarray}
the function $F(x)$ can be made to vanish by choosing for $\xi(x)$ any solution of $\xi_{xx}-\xi_x^2=F$.
A Lorentz transformation of the original null-tetrad, defined by
\begin{eqnarray}
\w{\omega}^1=\frac{1}{2}[-e^{i\pi/4}\w{\sigma}^1-e^{-i\pi/4}\w{\sigma}^2+i\w{\sigma}^3-i\w{\sigma}^4],\nonumber\\
\w{\omega}^3=\frac{1}{2}[e^{-i\pi/4}\w{\sigma}^1+e^{i\pi/4}\w{\sigma}^2+\w{\sigma}^3+\w{\sigma}^4],\nonumber\\
\w{\omega}^4=\frac{1}{2}[-e^{-i\pi/4}\w{\sigma}^1-e^{i\pi/4}\w{\sigma}^2+\w{\sigma}^3+\w{\sigma}^4],
\end{eqnarray}
allows one to write the line-element as $\ud s^2 = 2(\w{\sigma}^1\w{\sigma}^2-\w{\sigma}^3\w{\sigma}^4)$ with the following simple expressions for the basis one-forms:
\begin{eqnarray}
\w{\sigma}^1= \frac{1}{2\sqrt{2}}[(\mathcal{E}z^2-\frac{i}{m^2+r^2}\mathcal{E}^{-1})\ud x -\mathcal{E}\ud z],\nonumber\\
\w{\sigma}^3=\frac{1}{2(m^2+r^2)}[m \ud t+(m+r)z \ud x+r \ud y] ,\nonumber\\
\w{\sigma}^4=\frac{1}{2(m^2+r^2)}[m \ud t+(m-r)z \ud x-r \ud y].\label{negQ}
\end{eqnarray}

\section{Energy-momentum tensor}
We investigate whether there exist pure radiation, Einstein-Maxwell or perfect fluid space-times in the conformal classes with representants (\ref{posQcaseA}), (\ref{posQcaseB}) and (\ref{negQ}). The equations will be tackled in the Newman-Penrose formalism, fixing boost and rotation as in the previous section. Let us first outline the general scheme to be followed. The constant spin coefficients of the chosen tetrad ${\cal B}:=(m^a,\overline{m}^a, \ell^a, k^a) \leftrightarrow (\delta, \overline{\delta}, \Delta, D)$ in the original spacetimes are given by (\ref{spincoeffs}), $\tau=\pi=m$ and $\mu=Q\rho=i r$. Herewith the components of the trace-free  Ricci tensor, as calculated from the Newman-Penrose  equations, are
\begin{eqnarray}
\Phi_{rs} & = & \left(
\begin{array}{ccc}
r^2       & -5Qrm i                                          & -m^2\\
5Qrm i & \frac{7}{2}\left(Qr^2-m^2\right)  & 5mri\\
-m^2    & -5mri                                              & r^2
\end{array}\right),
\end{eqnarray}
while the Ricci scalar and the only non-zero Weyl scalar $\Psi_2$ are given by (\ref{KS3const}). As in the above discussion, we are interested in space-times with $m>0$, $r>0$ and $\Psi_2\neq 0$. We now perform a conformal transformation $ds^2\rightarrow ds'^2=\Omega^{-2}ds^2$, and take ${\cal B}':=( m'^a,\overline{m}'^a,\ell'^a,k'^a)=(\Omega\,m^a,\Omega\,\overline{m}^a,\Omega\,\ell^a,\Omega\,k^a)$ as the NP null tetrad for $ds'^2$. The spin coefficients of this tetrad are
\begin{eqnarray*}
\kappa'=\nu'=\sigma'=\lambda'=0,\quad \tau'=\Omega m+\delta\Omega,\quad \pi'=\Omega m-\bar{\delta}\Omega,\\
\beta'=\frac 32\Omega m-\frac 12\delta\Omega, \quad\alpha'=\frac 32\Omega m+\frac 12\bar{\delta}\Omega,\quad\rho'=i\Omega Qr+D\Omega,\\
\mu'=i\Omega r-\Delta\Omega,\quad  \epsilon'=\frac 32i\Omega Qr-\frac 12 D\Omega,\quad \gamma'=\frac 32i\Omega r+\frac 12\Delta\Omega,
\end{eqnarray*}
the appearing directional derivative operators still being the vectors of the tetrad ${\cal B}$.
Substituting this in the Newman-Penrose equations one obtains
\begin{equation}
\Psi_0'=\Psi_1'=\Psi_3'=\Psi_4'=0,\quad \Psi_2'=-\frac{8\Omega^2}{3}(m^2+Qr^2),
\end{equation}
in accordance with the conformal transformation properties of the Weyl tensor, and
\begin{subequations}\label{RicciConftran}
\begin{eqnarray}
\Phi_{00}' & = & \Omega^2r^2+\Omega D(D(\Omega)),\label{Phi00}\\
\Phi_{01}' & = & -5\Omega^2Qrm i-\Omega Qri\delta(\Omega)-3\Omega m D(\Omega)+\Omega\delta(D(\Omega)),\\
\Phi_{02}' & = & -\Omega^2m^2+\Omega\delta(\delta(\Omega)),\\
\Phi_{11}' & = &  \frac{\Omega}{2} \left(r i Q  \Delta(\Omega) +  r i D(\Omega)+ m \bar{\delta}(\Omega)+ m \delta(\Omega)\right)\\
                      &&\;\;\;\;\; +\frac{7}{2} \Omega^2 \left(Q r^2- m^2\right)+\frac{\Omega}{2}\Delta(D(\Omega))+\frac{\Omega}{2}\bar{\delta}(\delta(\Omega)),\nonumber\\
\Phi_{12}' & = &    5\Omega^2mr i+\Omega\delta(\Delta(\Omega))+3\Omega m\Delta(\Omega)-\Omega ri\delta(\Omega),\\
\Phi_{22}' & = &    \Omega ^2r^2+\Omega \Delta(\Delta(\Omega )),\\
R'               & = & 20 \Omega ^2 (Qr^2+m^2)+24 \Delta(\Omega ) D(\Omega )-24 \delta(\Omega ) \bar{\delta}(\Omega )\\
                       &&\;\;\;\;\; +6 \Omega  \left(\bar{\delta}(\delta(\Omega ))+ \delta(\bar{\delta}(\Omega )) -  \Delta(D(\Omega ))-D(\Delta(\Omega ))\right).\nonumber
\end{eqnarray}
\end{subequations}
Conditions on the energy-momentum of $ds'^2$ lead via Einstein's equations to conditions on $\Phi_{rs}'$ and $R'$. Equations (\ref{RicciConftran}) and their complex conjugates form 10 real PDEs, and together with the 6 NP commutator relations applied to $\Omega$, these allow one to solve for all second order derivatives of $\Omega$. The resulting integrability conditions, which are equivalent to the 20 NP Bianchi equations for $ds'^2$, form a set of first-order PDEs which will be analyzed in this section.

\subsection{Pure Radiation}
The space-time with metric $ds'^2$ is a pure radiation space-time iff its energy-momentum tensor is given by $T_{ab}=\Phi n_an_b$, where $n^a$ is a null vector,
\begin{equation}
n_1n_2-n_3n_4=0.\label{null}
\end{equation}
Using Einstein's equations with cosmological constant $\Lambda$, $G_{ab}=T_{ab}-\Lambda g_{ab}$, this translates to
\begin{eqnarray}
\Phi_{rs}' & = & \frac{\Phi}{2}\left(
\begin{array}{ccc}
n_4^2    & n_1n_4                                                      & n_1^2\\
n_2n_4 &  \frac{1}{2}\left(n_1n_2+n_3n_4\right) & n_1n_3\\
n_2^2    & n_2n_3                                                      & n_3^2
\end{array}\right),\label{purerad Phi}\\
R' & = & 4\Lambda.
\end{eqnarray}
After solving for all second order derivatives of $\Omega$ as discussed above, one should analyze the integrability conditions. For aligned solutions $n_1=n_2=0$, these imply
\begin{equation}\label{scalareq}
Qr^2+m^2=0.
\end{equation}
Hence, for $Q=1$ there are no solutions, whereas for $Q=-1$, the solutions are conformally flat.

For non-aligned solutions one can eliminate $n_2$ using (\ref{null}), and the integrability conditions yield linear first order differential equations for $n_1$, $n_3$, $n_4$, $\Omega$ and $\Phi$. Elimination of the first order derivatives yields the same scalar equation (\ref{scalareq}),
hence there are no non-conformally flat solutions.

\subsection{Einstein-Maxwell fields}

The gravitational field is an Einstein-Maxwell field iff with respect to ${\cal B}'$ one has
\begin{eqnarray}
\Phi_{rs}' & = & \left(
\begin{array}{ccc}
F_0\overline{F_0}    & F_0\overline{F_1}                                                      & F_0\overline{F_2}\\
F_1\overline{F_0} &  F_1\overline{F_1} & F_1\overline{F_2}\\
F_2\overline{F_0}   & F_2\overline{F_1}                                                      & F_2\overline{F_2}
\end{array}\right),\\
R' & = & 4\Lambda,
\end{eqnarray}
where $\Lambda$ is a possible cosmological constant, and the complex fields $F_0$, $F_1$ and $F_2$ moreover satisfy the Maxwell equations
\begin{eqnarray}
D(F_1)-\bar{\delta}(F_0) = F_1[2Qr i+D(\ln(\Omega^2))]-F_0[2m+\bar{\delta}(\ln(\Omega^2))],\label{DF1}\\
\Delta(F_1)-\delta(F_2)=F_1[-2r i+\Delta(\ln(\Omega^2))]+F_2[2m-{\delta}(\ln(\Omega^2))],\nonumber\\
\delta(F_1)-\delta(F_0)=F_1[2m+{\delta}(\ln(\Omega^2))]-F_0[2r i+\Delta(\ln(\Omega^2))],\nonumber\\
\bar{\delta}(F_1)-D(F_2)=F_1[-2m+\bar{\delta}(\ln(\Omega^2))]+F_2[2Qr i-D(\ln(\Omega^2))].\nonumber
\end{eqnarray}
As the trace-free Ricci tensor of a null Maxwell field ($F_0F_2=F_1^2$) has the algebraic structure (\ref{purerad Phi})~\cite{Kramer}, this case is excluded by the result of the previous paragraph. Regarding potential non-null fields, one can distinguish between the cases where at least one of its null eigendirections is aligned with a principal null direction of the Weyl tensor ($F_0=0\neq F_1$ and/or $F_2=0\neq F_1$) or not ($F_0F_2\neq 0$). In the latter case an overdetermined system of integrability conditions arises, but we have been unable to decide on the (non-)existence and ampleness of solutions. The aligned case turns out to be excluded: taking $\w{k}$ as an aligned null vector ($F_0=0\neq F_1$), and combining the $[\Delta,D]$  and $[\delta,\bar{\delta}]$ commutator relations applied to $D(\Omega)$ (also making use of (\ref{DF1}), $D(\Lambda)=0$ and $\Psi_2'\neq 0$) one finds $D(\Omega)=0$, in contradiction with (\ref{Phi00}). This confirms the statement in \cite{McLenVdB1993} that doubly
aligned Petrov type D electrovacs have conformal representants admitting a Killing spinor with non-constant eigenvalues.


\subsection{Perfect Fluid}
The space-time with metric $ds'^2$ is a perfect fluid (PF) space-time iff its energy-momentum is given by $T_{ab}=Su_au_b+pg_{ab}$, where $u^a$ is a unit timelike vector,
\begin{equation}
u_1u_2-u_3u_4=-\frac{1}{2}.\label{unitTime}
\end{equation}
Using Einstein's equations, this translates to
\begin{eqnarray}
\Phi_{rs}' & = & \frac{S}{2}\left(
\begin{array}{ccc}
u_4^2    & u_1u_4                                                      & u_1^2\\
u_2u_4 &  \frac{1}{2}\left(u_1u_2+u_3u_4\right) & u_1u_3\\
u_2^2    & u_2u_3                                                      & u_3^2
\end{array}\right),\\
R' & = & S-4p.\label{RicciPF}
\end{eqnarray}
One can again solve these equations together with the NP commutation relations for all second order derivative operators in $\Omega$, in terms of $p$, $S$, $u_a$, $\Omega$ and first order derivatives of $\Omega$.  For aligned solutions $u_1=u_2=0$, $u_3={1}/{(2u_4)}$ the integrability conditions imply $Qr^2+m^2 = 0$, such that there are no non-conformally flat solutions.

For non-aligned solutions, one can eliminate $u_2$ using equation (\ref{unitTime}). The integrability conditions form a set of first order PDEs for $\Omega$, $u_1$, $u_3$, $u_4$, $S$ and $p$.  Interpreting these as linear equations for the derivatives of the variables, these are consistent iff
\begin{eqnarray}
u_3     & = & Q u_4\label{equ3}\\
u_4^2 & = & u_1^2+1/2.\label{equ4}
\end{eqnarray}
$u_a$ cannot satisfy equation (\ref{unitTime}) if $u_3=-u_4$, as $u_2=\overline{u_1}$, which implies that there are no PF solutions for $Q=-1$. If on the other hand $Q=1$, substitution of equations (\ref{equ3}) and (\ref{equ4}) in the integrability conditions, yields a new set of differential equations linear in the derivatives of $\Omega , u_1,S,p$. Elimination of the first order derivatives yields the scalar equation
\be
u_1^2 = -\frac{1}{4}.
\ee
This is however in contradiction with equations (\ref{unitTime}), (\ref{equ3}), (\ref{equ4}), which imply $u_1=u_2=\overline{u_1}$. The constructed class of KS space-times therefore contains no PF solutions.

\section{Conclusion}
In \cite{KS1_2} it was left as an open question whether space-times which violate both equation (\ref{condKS1}) and equation (\ref{condKS2}) exist. We showed that the answer lies in the affirmative, and that the conformal representant with $|X|=1$ admits a 3- or 4-dimensional maximal group of isometries. The line-element for the latter space-times was constructed, and it was shown that their respective conformal classes do not contain any PF or pure radiation members. It remains an open question whether these classes contain a, necessarily non-aligned, Einstein-Maxwell space-time. This is an intriguing problem, as almost all Petrov type D Einstein-Maxwell solutions known to date belong to the aligned family. The integration of the space-times admitting a 3-dimensional maximal group of isometries will be discussed elsewhere.

\section*{Acknowledgement}

We thank Stefan Haesen for pointing out the coordinate transformation (\ref{Stefan}). D. B. is supported by the Research Foundation-Flanders (FWO). L.W. is supported by a BOF Research Grant (UGent) and a FWO mobility grant.

\section*{References}


\providecommand{\newblock}{}

\end{document}